\documentclass[journal,twoside,web]{ieeecolor}
\usepackage{jsen}
\usepackage{cite}
\usepackage{amsmath,amssymb,amsfonts}
\usepackage{algorithmic}
\usepackage{graphicx}
\usepackage{textcomp}
\usepackage{wrapfig}
\def\BibTeX{{\rm B\kern-.05em{\sc i\kern-.025em b}\kern-.08em
    T\kern-.1667em\lower.7ex\hbox{E}\kern-.125emX}}
\definecolor{abstractbg}{rgb}{0.89804,0.94510,0.83137}
\begin{document}
\title{Modeling and simulation of the R5912 photomultiplier for the LAGO project}
\author{J. Pe\~na-Rodr\'iguez, S. Hern\'andez-Barajas, Y. Le\'on-Carre\~no and L. A. N\'u\~nez
\thanks{The authors acknowledge financial support of the Halley Research Group at Universidad Industrial de Santander. We are also very thankful to LAGO and to the Pierre Auger Collaboration for their continuous support. }
\thanks{J. Pe\~na-Rodr\'iguez is with the Escuela de F\'isica, Universidad Industrial de Santander, Carrera 27 Calle 9, 640002 Bucaramanga, Colombia (e-mail: jesus.pena@correo.uis.edu.co). }
\thanks{S. Hern\'andez-Barajas and Y. Le\'on-Carre\~no were with the Escuela de Ingenierías El\'ectrica, Electr\'onica y de Telecomunicaciones, Universidad Industrial de Santander, Carrera 27 Calle 9, 640002 Bucaramanga, Colombia. (e-mail: sandra.hernandez4@correo.uis.edu.co and yesid.leon@correo.uis.edu.co).}
\thanks{L. A. N\'u\~nez is with the Escuela de F\'isica, Universidad Industrial de Santander, Carrera 27 Calle 9, 640002 Bucaramanga, Colombia (e-mail: lnunez@uis.edu.co).}}

\IEEEtitleabstractindextext{%
\fcolorbox{abstractbg}{abstractbg}{%
\begin{minipage}{\textwidth}%
% \begin{wrapfigure}[12]{r}{3in}%
% \includegraphics[width=3in]{jsenga.png}%
% \end{wrapfigure}%
\begin{abstract}
We present the results of modeling and simulating the Hamamatsu R5912 photomultiplier tube that is used in most of the sites of the Latin American Giant Observatory (LAGO). The model was compared with data of in-operation water Cherenkov detectors (WCD) installed at Bucaramanga-Colombia and Bariloche-Argentina. The LAGO project is an international experiment that spans across Latin America at different altitudes joining more than 35 institutions of 11 countries. It is mainly oriented to basic research on gamma-ray bursts and space weather phenomena. The LAGO network consists of single or small arrays of WCDs composed mainly by a photomultiplier tube and a readout electronics that acquires single-particle or extensive air shower events triggered by the interaction of cosmic rays with the Earth atmosphere.
\end{abstract}

\end{minipage}}}

\maketitle

\section{Introduction}
\label{sec:introduction}
\IEEEPARstart{A}{strophysical} phenomena are studied by means of giant cosmic ray (CR) observatories spread around the world. Such experiments, located at ground level, detects atmospheric particle showers resulting from the interaction of high energy primary CRs with atmospheric gases. The extensive air shower (EAS) detection is made using different techniques, taking advantage of the signal that charged particles leave along their pathway. At ground level the EAS is detected by arrays of Cherenkov counters, scintillators or antennas getting information of the shower front, composition and primary energy \cite{Bertou2006, Galindo2017}. The longitudinal development of the EAS is directly interpreted from the electromagnetic radiation created by photons, electrons and positrons crossing the atmosphere. This EAS component is detected by fluorescence telescopes \cite{Neesal2011}, Imaging Atmospheric (or Air) Cherenkov Telescopes \cite{Schoorlemmer2019} and radio antennas \cite{Aab2016, Schrder2016}. 

The LAGO project was founded with the goal of creating a collaborative project in astroparticle physics research to train young scientists in Latin America. LAGO consists of a network of own made water Cherenkov detectors (WCD) spanning over different sites, located at several latitudes (from Mexico to the Antarctic) and altitudes (from sea level up to 5000 m a.s.l.) \cite{Sidelnik2016}. 

The WCD network of LAGO is able to detect short duration transients --like gamma-ray bursts (GRBs)-- and long duration transients --like Forbush decreases-- \cite{Durn2016,Asorey2018} by searching changes in the cosmic ray background recorded using the single particle technique \cite{Vernetto2000, Leon2018}. LAGO operates in energies ranging from 0.5 GeV to tens of TeV.

LAGO detectors are made up of cylindrical containers of plastic, metal or fibreglass with an internal Tyvek coating for enhancing its optical properties (reflection and diffusion) and the transmission efficiency of Cherenkov photons generated by crossing charged particles. The Cherenkov radiation is usually collected by an 8$^{\prime\prime}$ Hamamatsu R5912 photomultiplier tube (PMT) located at the center of the WCD cover. The pulses generated by the anode and last dynode of the PMT are digitized by a 10-bit fast analog-to-digital converter working at 40\,MHz. A 12-sample records the pulses with a 25 ns resolution timestamp.

A key point in the LAGO WCDs is the calibration process. We establish a conversion rule from the digitized charge in electronic units to deposited energy in vertical-equivalent muons (VEM) \cite{Bertou2006, Galindo2017}. This relationship depends on the linearity of both the PMT and the readout electronics, working together. 

This paper propose a general PMT and bias chain model which is tuned with the LAGO's current PMT parameters. The model allow us to assess the electronics front-end linearity under different acquisition conditions. The SPICE simulation performance is validated with data measurements from the Nahuelito, Chitaga, and MuTe detectors.

\section{Methods}

\subsection{A generic PMT model}

\begin{figure}[h!]
\begin{center}
\includegraphics[width=0.48\textwidth]{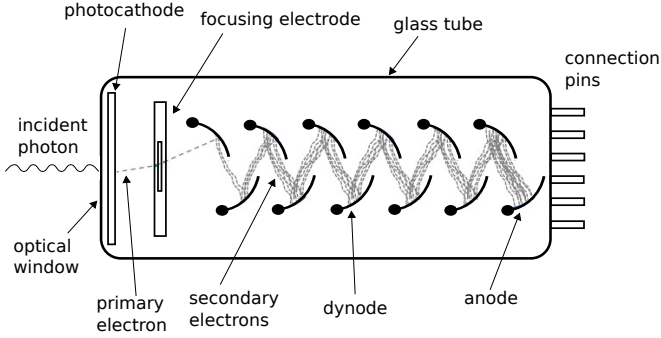}
\caption{PMT functioning sketch. The incident photon impinges the photocathode releasing a primary electron which create a secondary electron avalanche due to the electric field generated along the dynodes. All the PMT parts are encapsulated in a vacuum glass tube.}
\label{PMT_sketch}
\end{center}
\end{figure}

A PMT is an optoelectronic device which generates a measurable electric current ($\sim$ mA) by means of the photoelectric effect when a photon impinges its photocathode. The photoelectron is accelerated by a potential difference reaching the energy for pulling up more electrons from the next dynode. This avalanche of secondary electrons along the dynodes amplifies the anode current with gain factors of $\sim10^6$-$10^7$. (See Fig. \ref{PMT_sketch}).

We modeled the PMT R5912 taking into account such basic principle of functioning and its intrinsic parameters: the number of amplification stages and the gain curve. The total gain of the PMT model is defined as,

\begin{equation}
G = \frac{I_a}{I_k},
\end{equation}

\noindent where $I_a$ is the anode current and $I_k$ is the photocathode current.

The PMT gain can be expressed as a function of the gain in each stage,

\begin{equation}
G =  \beta \prod_{i=1}^{N}  g_i ,
\label{eq::totalgain}
\end{equation}

\noindent where $g_i$ is the gain in each stage, $N$ is the number of dynodes and $\beta$ is the collection efficiency. The gain $g_i$ depends on the inter-dynode voltage $v_i$,

\begin{equation}
g_i = k_i v_i^\alpha ,
\end{equation}

\noindent where $k_i$ is a constant and $ 0.6  \leq \alpha \leq 0.8$ is an intrinsic parameter of the PMT. The total gain \eqref{eq::totalgain} can be expressed as the product of all the inter-dynode gains or in function of the PMT bias voltage $V_B$,

\begin{equation}
G =  \prod_{i=1}^{N}  k_i (V_B \epsilon_i)^{\alpha} ,
\label{eq::gain}
\end{equation}

\noindent where $\epsilon_i$ is the fraction of the bias voltage in each inter-dynode stage as a result of the resistor polarization chain.

The fraction of the bias voltage is defined as 
\begin{equation}
\epsilon_i = \frac{R_i}{R_T} ,
\end{equation}

\noindent where $R_i$ is the interdynode resistance and $R_T$ is the total resistance of the polarization chain.

To simplify the model, we can assume that $k_i$ values are equal for all dynodes due it depends on the dynode material \cite{Akimov2017}. Equation \ref{eq::gain} is transformed in

\begin{equation}
G =  k^N V_B^{N\alpha} \left ( \prod_{i=1}^{N} \epsilon_i \right)^{\alpha} .
\label{eq::gainkte}
\end{equation}

We define $\varepsilon$, to estimate the value of $\alpha$ and $k$, as

\begin{equation}
    \varepsilon = \sqrt[N]{\prod_{i=1}^{N} \epsilon_i} .
    \label{eq::varep}
\end{equation}

Replacing \eqref{eq::varep} in \eqref{eq::gainkte}, we expressed the gain as follows,

\begin{equation}
G = k^N (V_B \varepsilon)^{N\alpha} . 
\label{eq::gainmodel}
\end{equation}

\subsection{Modeling the R5912 PMT}

\begin{figure}[h!]
\begin{center}
\includegraphics[width=0.4\textwidth]{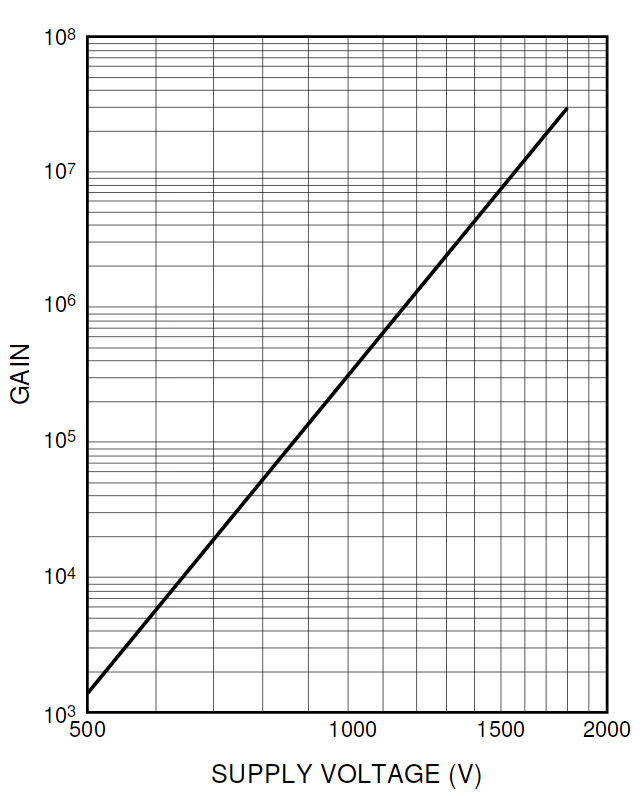}
\caption{Gain curve of the R5912 PMT. The gain of the PMT has an exponential relation depending on the high voltage applied between the anode and cathode \cite{Hamamatsu2019}.}
\label{Gain_curve}
\end{center}
\end{figure}

To get the parameters $\alpha$ and $k$, a couple of points $[V_{B1}, G_1]$ and $[V_{B2}, G_2]$ are extracted from the gain curve of the Hamamatsu R5912 PMT\cite{Hamamatsu2019}. (See Fig. \ref{Gain_curve}).

The values [1000\,V, 3$\times 10^5$] and [1500\,V, 7$\times 10^6$] were chosen. We derived a pair of equations from \eqref{eq::gainmodel} with the given points to solve the unknown variables ($\alpha$, $k$).

\begin{equation}
G_{1} = k^N (V_{B1} \varepsilon)^{N \alpha},
\label{eq:gain1}
\end{equation}

\begin{equation}
G_{2} = k^N (V_{B2} \varepsilon)^{N \alpha},
\label{eq:gain2}
\end{equation}

\noindent where the number of dynodes is $N=10$. The parameter $\varepsilon$ is calculated by means of the voltage distribution ratio in the resistive polarization chain, provided in the PMT datasheet, as shown in Table \ref{net}.

\begin{equation}
\varepsilon = 0.035 .
\end{equation}

\begin{table}[ht]
\centering
  \caption{Tapered voltage distribution of the PMT R5912 for linear measurements \cite{Hamamatsu2019}.}
  \begin{tabular}{ | c | c | c |}
    \hline
    Electrodes & $R_i$ & $\epsilon_i$ \\ \hline
    K-Dy1 & 11.3 & 0.308 \\ \hline
    Dy1-F2 & 0 & 0 \\ \hline
    F2-F1 &  0.6 & 0.016 \\ \hline
    F1-F3 & 0 & 0 \\ \hline
    F3-Dy2 & 3.4 & 0.092 \\ \hline
    Dy2-Dy3 & 5 & 0.136 \\ \hline
    Dy3-Dy4 & 3.33 & 0.090 \\ \hline
    Dy4-Dy5 & 1.67 & 0.045 \\ \hline
    Dy5-Dy6 & 1 & 0.027 \\ \hline
    Dy6-Dy7 & 1.2 & 0.032 \\ \hline
    Dy7-Dy8 & 1.5 & 0.040 \\ \hline
    Dy8-Dy9 & 2.2 & 0.060 \\ \hline
    Dy9-Dy10 & 3 & 0.081 \\ \hline
    Dy10-P & 2.4 & 0.065 \\
    \hline
  \end{tabular}
  \label{net}
\end{table}

Then, an expression for $k$ is obtained from \eqref{eq:gain2} as follows,

\begin{equation}
k=\sqrt[N]{\frac{G_2}{(V_{B2}\varepsilon)^{N \alpha}}} ,
\label{eq:k0}
\end{equation}
\noindent and replacing \eqref{eq:k0} in \eqref{eq:gain1} the parameter $\alpha$ is,

\begin{equation}
\alpha=\frac{\log \left( \frac{G_1}{G_2} \right)}{N \log \left( \frac{V_{B1}}{V_{B2}} \right)} .
\label{eq:alpha}
\end{equation}
From \eqref{eq:k0} and \eqref{eq:alpha} we obtain $k=0.223$ and $\alpha= 0.776$.

\subsection{PMT and passive biasing network Spice simulation}

The dynodes and anode currents were modeled as function of the parameters $k$, $\alpha$ , $\epsilon_i$, $V_B$, and $N$. The current flowing through $i$th dynode is defined as,

\begin{equation}
I_{d,i} = I_k \frac{(k V_B^\alpha)^N \left ( \prod_{i=1}^{N} \epsilon_i \right)^{\alpha}}{(k v_i^\alpha)^{N+1-i} \left ( \prod_{i=1}^{N+1-i} \epsilon_i \right)^{\alpha} }, \ \ \ i=1,2, \cdots N .
\label{Id}
\end{equation}

The anode current is,
\begin{equation}
I_a = I_k k^N (V_B \varepsilon)^{N \alpha} .
\label{Ia}
\end{equation}

The PMT and the biasing network were simulated using the Orcad Pspice software. We used the GVALUE block to model the PMT currents flowing from the cathode to the anode along each PMT dynode \cite{Krihely2014}. This block sets the transfer function described by \eqref{Id} and \eqref{Ia} for each amplification stage depending on the voltage applied between adjacent dynodes. 

Resistive divider networks are the most widely used method to bias PMTs. We selected a tapered resistive chain with decoupling capacitors to reduce nonlinearities in the PMT response due to space-charge effect (large current flowing in the dynodes) in pulse-mode operation \cite{Huang_2013, Hamamatsu2007}. The resistor values were estimated taking into account the interdynode ratios presented in the Table \ref{net}. Decoupling capacitors of 20 nF were connected (serial and parallel) in the last six dynodes and the anode.

\begin{figure}[h!]
\begin{center}
\includegraphics[width=0.45\textwidth]{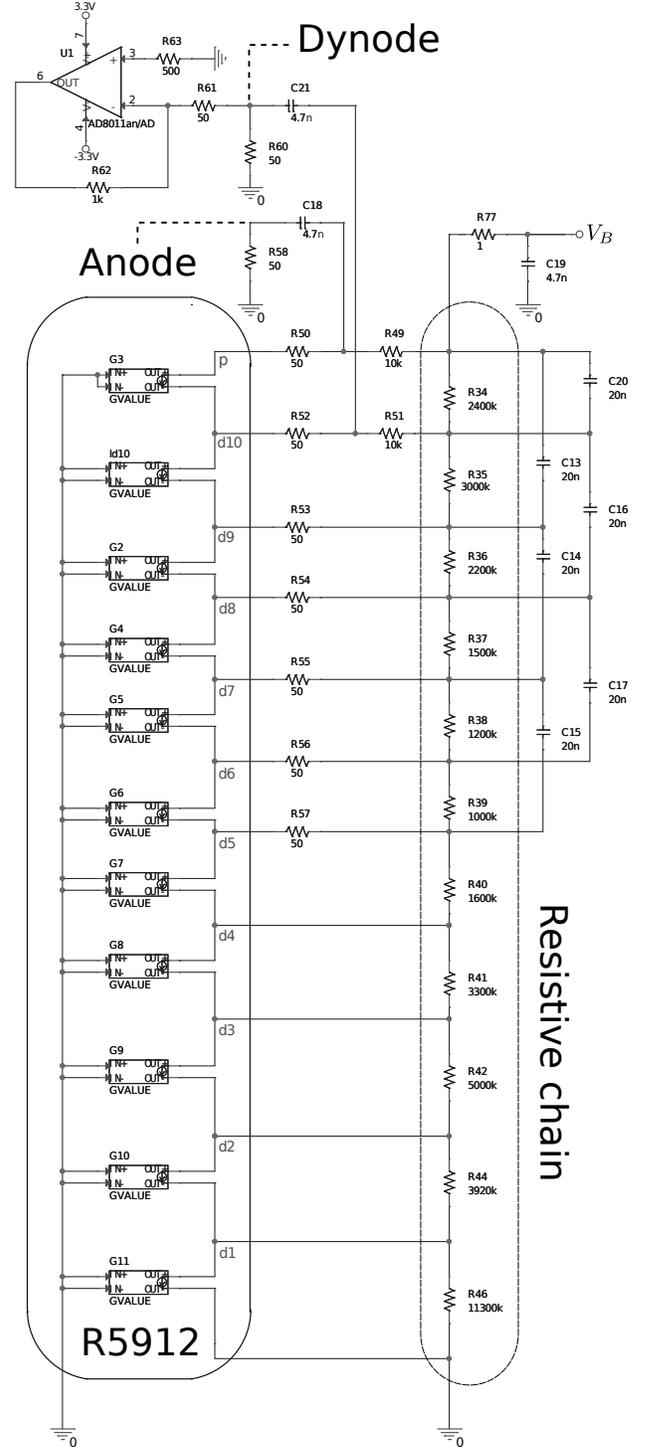}
\caption{Spice model of the R5912 PMT and the tapered resistive chain.}
\label{Circuit}
\end{center}
\end{figure}

The PMT output signal has a high direct current (DC) bias which can destroy the frontend electronics. We install coupling capacitors of 4.7\,nF (C18 and C21) to filter the DC component in the anode and the last dynode output. For avoiding oscillations in the signal due to reflections for bad impedance coupling in the transmission lines we implemented 50\,$\Omega$ output loads. 

An amplification stage was connected to the last dynode output to increase the dynamic response/range if the dynode pulse amplitude saturates the readout system we can recover the pulse shape from the anode output. The operational amplifier AD8011 amplifies 20 times the dynode output and inverts its polarity. The Fig. \ref{Circuit} shows the schema of the designed Spice model.

\subsection{Incident photon yield and cathode current}

We carried out simulations using the particle-matter interaction code GEANT4 to characterize the mean incident photon signal on the PMT cathode generated by charged particles crossing the WCD. We injected 10$^5$ muons of 3\,GeV perpendicularly to a 120\,cm height WCD \cite{VsquezRamrez2020, Caldern2015}. The average number of Cherenkov photons ($N_{\gamma}$) along the path were 46857, 1617 of such photons reach the PMT optical window and the PMT photocathode releases around 203 photo-electrons ($N_{pe} = \eta N_{\gamma}$) taking into account the maximum quantum efficiency ($\eta = 22 \%$ at 390\,nm). 

\begin{figure}[h!]
\begin{center}
\includegraphics[width=0.45\textwidth]{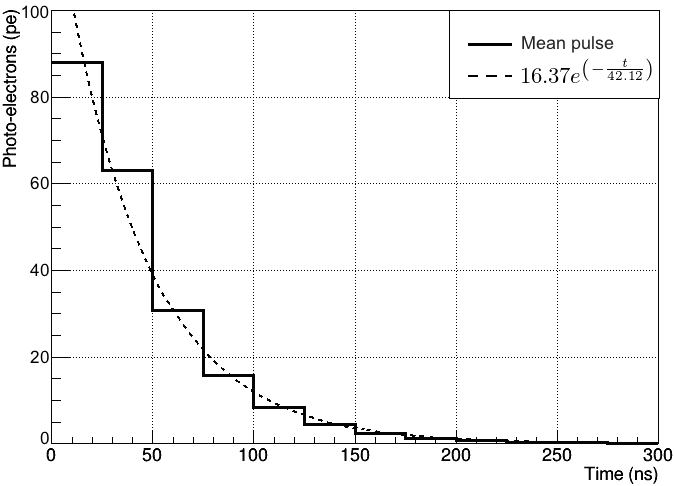}
\caption{Number of Cherenkov photons impinging the PMT for 3\,GeV muons crossing the WCD. The solid-line represents the average number of photo-electrons and the dashed-line the best exponential fit. The attenuation time is $\sim$ 42.12\,ns and the pulse width (at the 10$\%$ amplitude) is $\sim$100\,ns \cite{VsquezRamrez2020}}
\label{pulse_G4}
\end{center}
\end{figure}

The shape of the photoelectron pulse at the PMT photocathode depends on the arrival time of the incident photons as shown in Fig. \ref{pulse_G4}. The pulse decreases exponentially having an time constant of $\sim$42.12\,ns and a time width (at the 10$\%$ amplitude) of $\sim$100\,ns.

The photocathode current $I_k$ is, 

\begin{equation}
I_k = \frac{Q}{t},
\end{equation}

\noindent where $Q$ is the electric charge in the photocathode,

\begin{equation}
Q = N_{pe}*e ,
\end{equation}
\noindent with $e$ the electron charge ($1.6 \times 10^{-19}$\,C). 

\begin{figure}[h!]
\begin{center}
\includegraphics[width=0.45\textwidth]{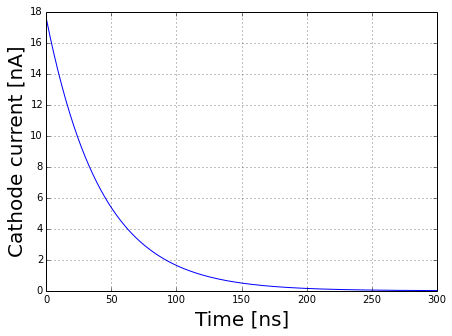}
\caption{Photocathode current taking into account the PMT quantum efficiency ($\eta = $ 22 $\%$) and the number of Cherenkov photons created by 3\,GeV muons crossing the WCD.}
\label{cathode}
\end{center}
\end{figure}

In Fig. \ref{cathode} we show the estimated photocathode current for 3\,GeV vertical muons impinging the WCD. The maximum peak of the current is $\sim$17 nA which can generate a 17\,mA anode current when the PMT gain is 10$^6$. The maximum anode dark current (unwanted current which occurs even in the absence of incident light, resulting from thermally excited electrons ) establishes the low boundary of acquisition 0.7\,$\mu$A.

\section{RESULTS}

\subsection{Simulated vertical muon charge}

The PMT model was biased at 1000\,V (2.9$\times$10$^5$ gain). When a vertical muon hits the WCD, a current signal of $\sim$5\,mA is measured at the anode and a voltage pulse of 250\,mV appears across the load resistance (50\,$\Omega$). 

The LAGO readout system digitizes the PMT pulses at 40\,MHz with a resolution of 10 bits (1\,mV/UADC); the pulse shape is stored in a 12 samples vector (300\,ns) \cite{SofoHaro2016}. We emulate the digitization process of the model outputs to compare simulations and data. The resulting pulse charge of the simulated vertical muon was 321.6\,UADC differing in about 4$\%$ of the value obtained by the MuTe WCD (333\,UADC).

\subsection{Response of the PMT and bias chain model}

\begin{figure}[h!]
\begin{center}
\includegraphics[width=0.48\textwidth]{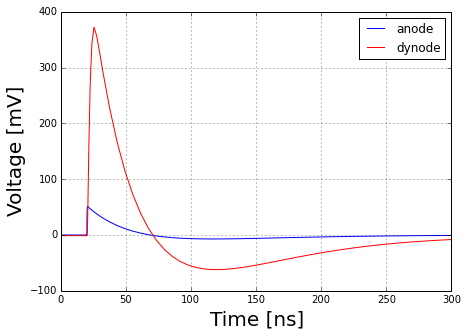}
\caption{Anode (blue) and dynode (red) outputs obtained from the Spice model at 1000\,V for a cathode current of 3.5\,nA. The dynode/anode ratios is 7.5 and the pulse width is $\sim$50\,ns.}
\label{Pulse}
\end{center}
\end{figure}

Fig. \ref{Pulse} shows the dynode and anode output for a photocathode current of 3.5\,nA and a bias voltage of 1000\,V. The dynode pulse maximum is 375\,mV and the anode is 50\,mV. The dynode/anode ratio  is $\sim$7.5 showing that the PMT amplifies 2.66 times the current flowing from the last dynode to the anode. The resulting pulse width $\sim$50\,ns occurs by action of the coupling capacitors (C18 and C21). The PMT transit time is not taken into account in the model.

\begin{figure}[h!]
\begin{center}
\includegraphics[width=0.4\textwidth]{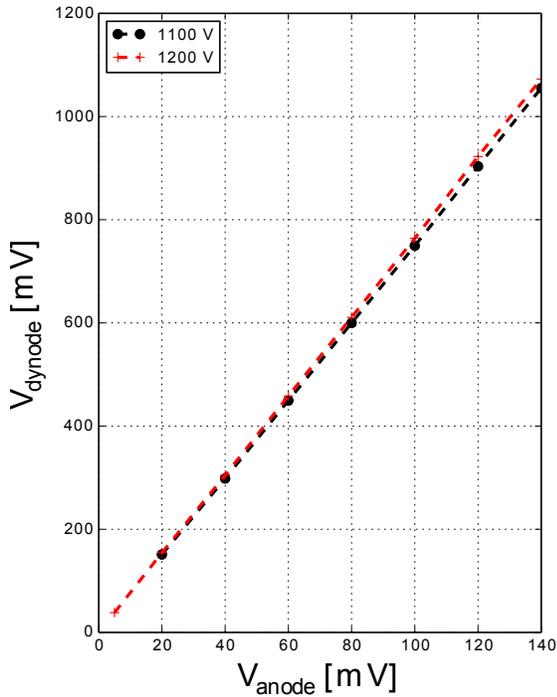}
\caption{Correlation between the dynode and anode output voltage at 1200\,V (red) and 1100\,V (black) for photocathode currents ranging 0.6-4.5\,nA.}
\label{Linear}
\end{center}
\end{figure}

The PMT and electronics readout must have a linear behaviour to guarantee an accurate estimation of the deposited energy of particles crossing the WCD. The linearity of the model was estimated correlating the dynode and anode pulse amplitude for different photocathode currents and bias voltages \cite{Arnaldi2019}. Fig. \ref{Linear} correlates the anode and dynode amplitudes for photocathode currents ranging between 0.6-2.2\,nA ($V_B$= 1200\,V) and 1.3-4.5\,nA ($V_B$= 1100\,V).

The curve slope increases sightly with the bias voltage from 7.52 at 1100\,V to 7.68 at 1200\,V representing a gain increment of $\sim$2. The linear response of the PMT breaks when the PMT reaches its electrical limits at 1800\,V (3$\times10^7$ gain) causing a saturation effect in the pulse amplitude.

\subsection{Model and data comparison}

\begin{figure}[h!]
\begin{center}
\includegraphics[width=0.4\textwidth]{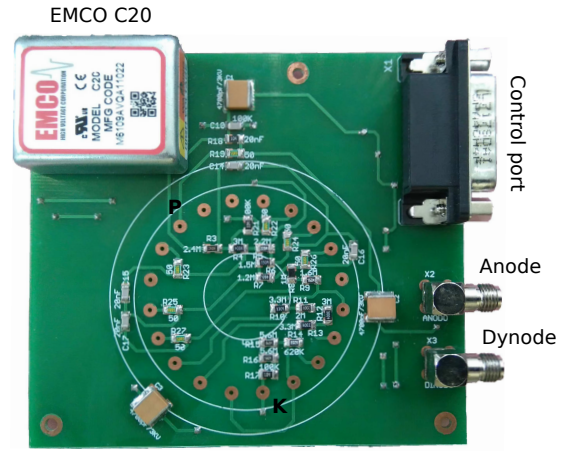}
\caption{PCB implementation of the Spice model. The bias voltage is supplied by an EMCO C20 DC/DC converter. The anode and dynode outputs are connected through 50\,$\Omega$ SMA connectors. A DB15 connector inputs the C20 control signal and the conditioning circuit (dynode amplification) supply. The tapered resistance chain was installed in the top layer while the conditioning circuit is in the bottom layer. The anode (P) and the cathode (K) electrodes are highlighted on the figure.}
\label{Base}
\end{center}
\end{figure}

We assess the model performance in two ways: a functional comparison with the present PMT base of LAGO, designed by the Pierre Auger Collaboration (Base-II) \cite{Genolini2001}, and a linearity comparison with data collected by the WCD Chitaga and Nahuelito. 

The PCB (Printed Circuit Board) of the proposed bias circuit (Base-I) is shown in Fig. \ref{Base}. The tapered resistive chain is biased by the EMCO C20 DC/DC converter. The output DC coupling was set by SMD (surface-mount device) capacitors to avoid electrostatic discharges and mechanical damages, as observed in the Base-II. The PCB was electrically isolated with a paint coating with a dielectric strength of 100\,kV/mm.

\begin{figure}[h!]
\begin{center}
\includegraphics[width=0.45\textwidth]{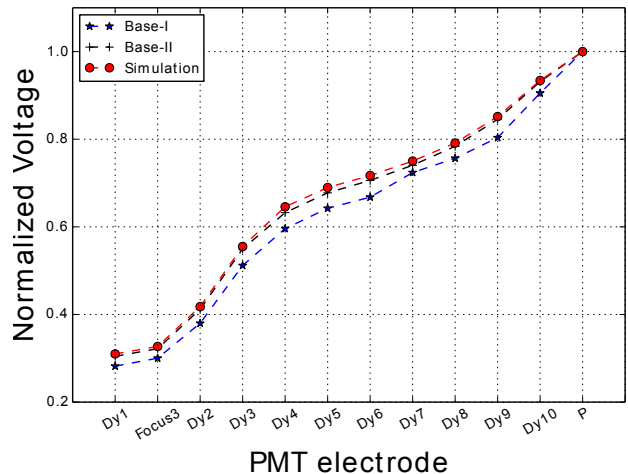}
\caption{Comparison of the electrode voltage distribution between the Spice model (red-line), the Base-I (blue-line) and II (black-line).}
\label{Comparison}
\end{center}
\end{figure}

The first test consisted of comparing the electrode voltage distribution of the Spice model and the Bases I and II. The data was normalized respect to the anode (P) voltage. From Fig. \ref{Comparison} we observe an average variation of 0.7$\%$ between the Base-II and the model while between the Base-II and the Base-I the variation is 2.8$\%$.

\begin{figure}[p]
\begin{center}
\includegraphics[width=0.45\textwidth]{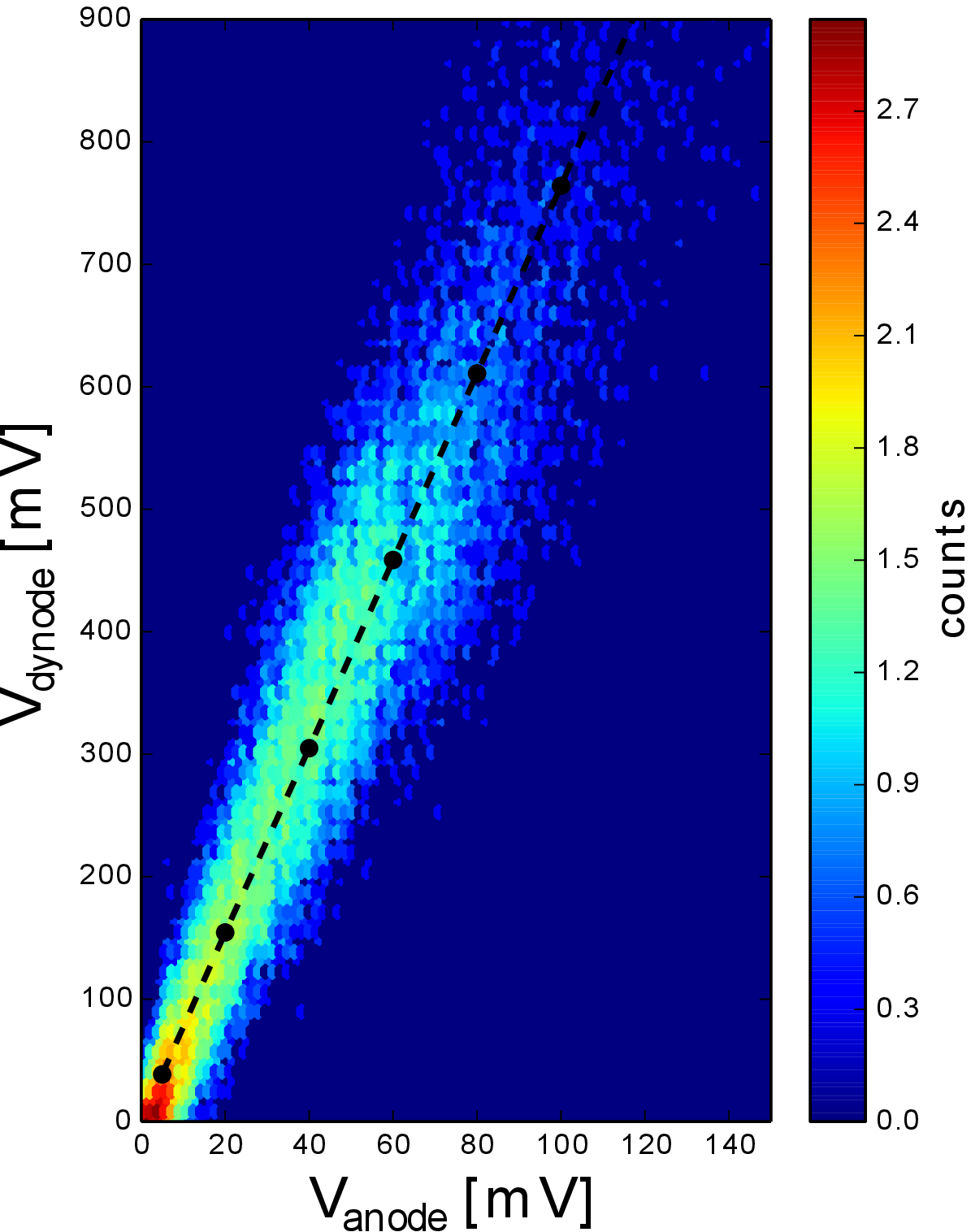}
\includegraphics[width=0.45\textwidth]{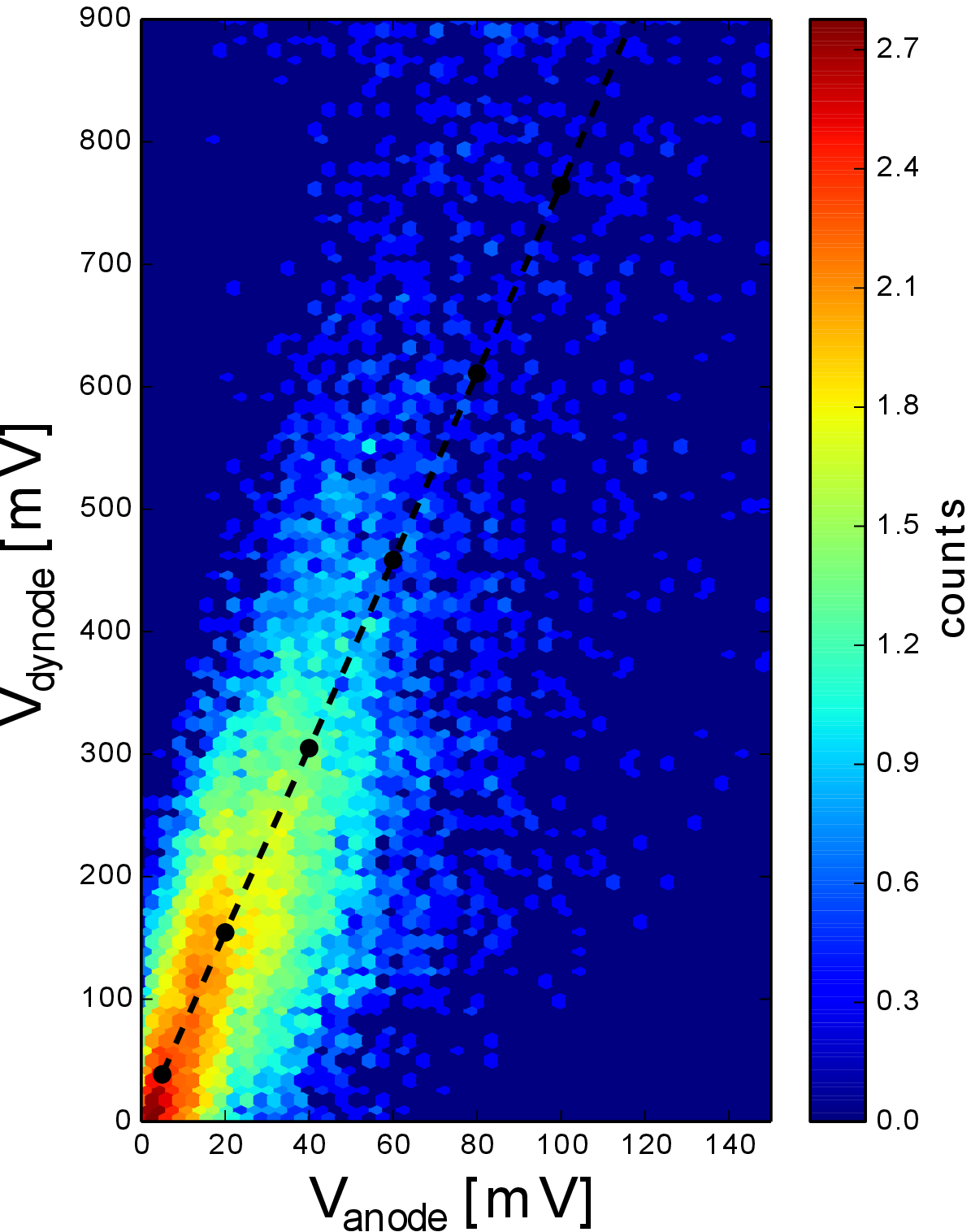}
\caption{Linearity measured on the WCDs Nahuelito and Chitaga operating at 1500~V and 1000~V respectively. The data shows the correlation between the maximum amplitude measured on the dynode and the anode with the LAGO's readout electronics. The dashed-line represents the model response taking into account the WCD operation conditions.}
\label{Nahuelito_Chitaga}
\end{center}
\end{figure}

The model was also compared with data of 30$\times 10^3$ pulses recorded by the Nahuelito and Chitaga WCDs as shown in Fig. \ref{Nahuelito_Chitaga}. The Nahuelito's PMT operates at 1500\,V with a discrimination threshold of 70 mV. The WCD data follows a linear distribution with the majority of the recorded events under 100 mV amplitudes. The dashed black-line represents the PMT response obtained from the Spice model.

The Chitaga's PMT operates at 1000~V with a discrimination threshold of 100 mV. The pulse charge distribution is linear but wider than Nahuelito because of the detector geometry differences.

\section{CONCLUSION}

A PMT and resistive chain model was designed and tested for the LAGO collaboration. The PMT model reproduces the expected gain depending on the bias voltage as well as the voltage distribution along the dynodes with a variance of $\sim$2.8$\%$. The model can be adapted to any PMT architecture by changing the number of electrodes, the voltage distribution ratio and the parameters $k$ and $\alpha$ -- derived from the PMT datasheet.

The vertical muon charge estimated by the model (321.6 UADC) differs only in 4$\%$ from the measured by the MuTe WCD (333 UADC). The linear correlation between the anode and dynode amplitudes of the model and the data recorded by the WCD Chitaga and Nahuelito were evaluated. 

In this PMT Spice model we set a uniform PMT quantum efficiency of 22$\%$ --the maximum. To obtain more accurate results, we recommend to use the quantum efficiency curve of the modeled PMT where the detection efficiency will change depending on the incident photon wavelength.

\appendices

%\begin{thebibliography}
\bibliographystyle{IEEEtran}
\bibliography{LAGO.bib}
%\end{thebibliography}

\vspace{3cm}

\begin{IEEEbiography}[{\includegraphics[width=1in,height=1.25in,clip,keepaspectratio]{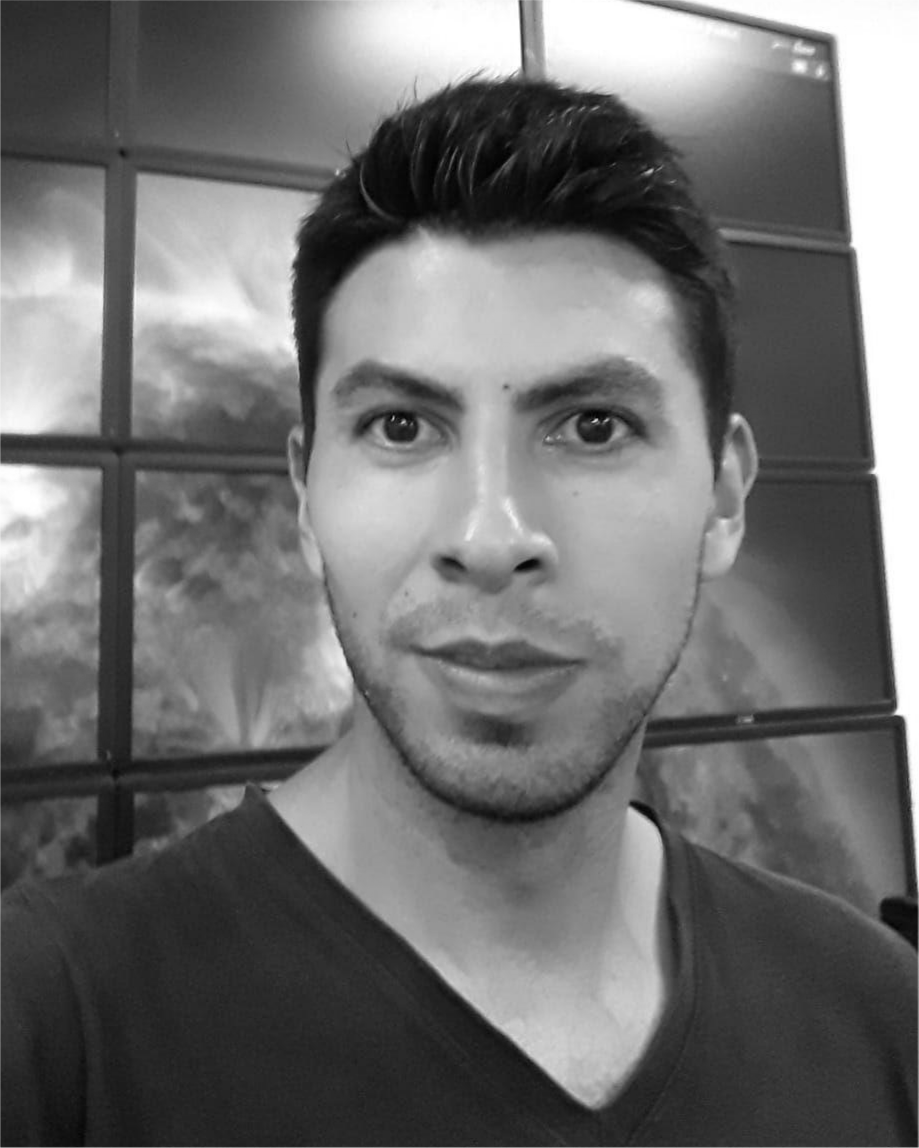}}]{J. Pe\~na-Rodr\'iguez} was born in Chitaga, Norte de Santander, Colombia in 1988. He received the B.S. degree in electronics engineering from the Universidad de Pamplona in 2011 and the M.S. degree in electronics engineering from the Universidad Industrial de Santander in 2016. He is currently pursuing a Ph.D. degree in physics at Universidad Industrial de Santander.

From 2014 to 2020, he was a Research Assistant with the Grupo Halley in the particle detector instrumentation area applied to space weather and muography. His research interests include particle physics and applications, machine learning, lightning detection, and weather monitoring. He takes part in the LAGO collaboration and the Pierre Auger Observatory.

\end{IEEEbiography}

\begin{IEEEbiography}[{\includegraphics[width=1in,height=1.25in,clip,keepaspectratio]{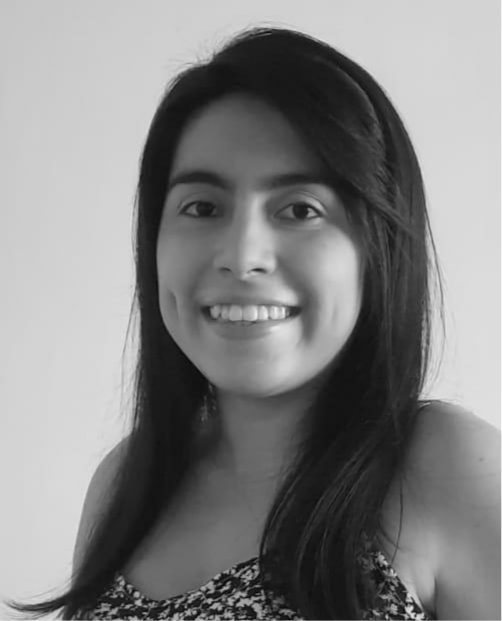}}]{S. Hern\'andez-Barajas} received the B.S. degree in electronics engineering from the Universidad Industrial de Santander in 2018. She worked in the assembly and calibration of the water Cherenkov detectors of the Guane array at Bucaramanga-Colombia. Her research interests include industrial instrumentation and control systems.

\end{IEEEbiography}

\begin{IEEEbiography}[{\includegraphics[width=1in,height=1.25in,clip,keepaspectratio]{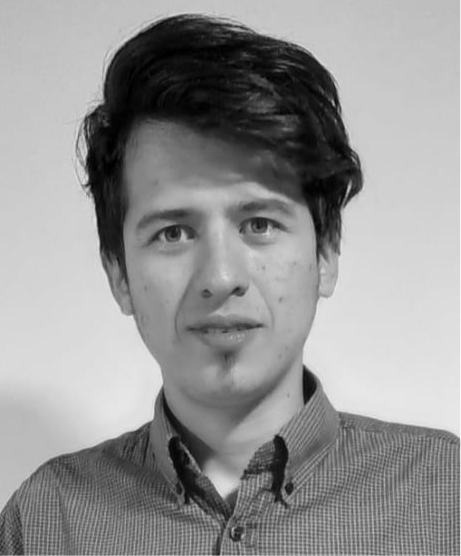}}]{Y. Le\'on-Carre\~no} received the B.S. degree in electronics engineering from the Universidad Industrial de Santander in 2018. He worked in the installation and calibration of the LAGO's water Cherenkov detectors at Bucaramanga-Colombia. He is interested in control systems and industrial automation.
\end{IEEEbiography}

\begin{IEEEbiography}[{\includegraphics[width=1in,height=1.25in,clip,keepaspectratio]{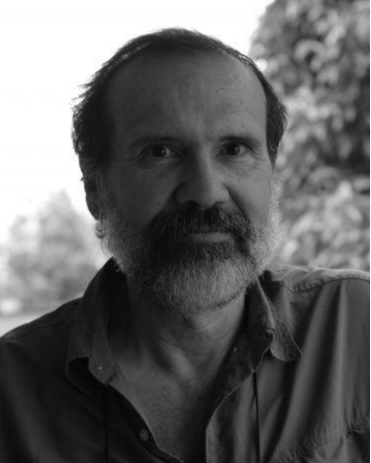}}]{L. A. N\'u\~nez} is a Senior Professor at Universidad Industrial de Santander, Bucaramanga-Colombia since 2010. From 1979 to 2009 he was a Former Senior Professor at Universidad de Los Andes, Mérida-Venezuela. Director of Information Technologies at the Universidad de los Andes from 1995 to 2009, and Director of the Venezuelan National Center for Scientific Computing (CeCalCULA) from 1999 to 2009. He has been involved in several European cooperation projects with Latin America.

He presently is a member of the Latin American Giant Observatory Collaboration (LAGO), the Pierre Auger Observatory Collaboration, and the Cooperación Latinoamericana de Redes Avanzadas (RedCLARA). His areas of interest are astroparticle physics, relativistic astrophysics, and information technologies.
\end{IEEEbiography}

\end{document}